\documentclass[pra,english,showpacs,preprintnumbers,amsmath,amssymb,nofootinbib,twocolumn,superscriptaddress]{revtex4-1}

\pdfoutput=1

\usepackage[latin1]{inputenc}
\usepackage{graphicx}
\usepackage{color}
\usepackage{bbm}
\usepackage{amssymb}
\usepackage{amsmath}
\usepackage{tabularx}

\usepackage{dsfont}

\def\0#1#2{\frac{#1}{#2}}

\def\s0#1#2{\mbox{\small{$ \frac{#1}{#2} $}}}

\newcommand{\beq}{\begin{equation}}
\newcommand{\eeq}{\end{equation}}
\newcommand{\bea}{\begin{eqnarray}}
\newcommand{\eea}{\end{eqnarray}}

\usepackage{babel}
\makeatother
\begin{document}

\title{Universality in one-dimensional fermions at finite temperature: \\
Density, pressure, compressibility, and contact}

\author{M. D. Hoffman} 
\author{P. D. Javernick} 
\author{A. C. Loheac} 
\author{W. J. Porter}
\author{E. R. Anderson} 
\author{J. E. Drut}

\affiliation{Department of Physics and Astronomy, University of North Carolina, Chapel Hill, North Carolina 27599-3255, USA}

\begin{abstract}
We present finite-temperature, lattice Monte Carlo calculations of the particle number density, compressibility, pressure, and 
Tan's contact of an unpolarized system of short-range, attractively interacting spin-1/2 fermions in one spatial dimension, 
i.e., the Gaudin-Yang model. In addition, we compute the second-order virial coefficients for the pressure and the 
contact, both of which are in excellent agreement with the lattice results in the low-fugacity regime.
Our calculations yield universal predictions for ultracold atomic systems with broad resonances in highly constrained traps.
We cover a wide range of couplings and temperatures and find results that support the existence of a strong-coupling
regime in which the thermodynamics of the system is markedly different from the non-interacting case. We compare
and contrast our results with identical systems in higher dimensions.
\end{abstract}


\maketitle
\section{Introduction} 
Universal aspects of strongly coupled nonrelativistic many-body systems have been in the spotlight for the last decade.
The realization and manipulation of these systems in the form of ultracold atomic clouds close to broad 
Feshbach resonances~\cite{RevExp}, followed by the enhanced 
understanding of their universality in terms of underlying conformal invariance, equations of state, and the 
Tan relations~\cite{RevTheory}, have clarified the central role of these simple systems for many-body quantum mechanics 
across all of physics. Broad resonances in dilute gases result in effective short-range interactions, such that 
the thermodynamics is universal~\cite{Universality}, in the sense that the only significant dynamical scale is the $s$-wave 
scattering length, and the thermal behavior is otherwise insensitive to the microscopic details of the system.

Interest in the one-dimensional (1D) version of these systems has existed in the area of condensed matter for a 
long time (see, e.g.~\cite{Giamarchi}), as many of these systems display quantum phase transitions, conformal invariance,
and in some cases are exactly solvable (at zero temperature).  Remarkably, 1D problems have also been studied in nuclear physics, where model calculations that resemble nuclear systems have 
often been performed (see, e.g.,~\cite{NegeleAlexandrouModel, Alexandrou}), both for insight into the physics as well 
as to develop new many-body methods~\cite{JurgensonFurnstahl}.

In spite of such broad interest, a precise characterization of unpolarized attractively interacting fermions in 1D
(e.g., in terms of the thermal equation of state and the contact) remains surprisingly absent from the literature.
Such a characterization is simultaneously a prediction for ultracold-atom experiments and a benchmark for many-body methods.
In contrast, there exists a considerable body of literature related to {\it polarized} Fermi gases in 1D, which are
particularly interesting in connection with exotic superfluid phases that may appear at low temperatures. Most of that 
work focuses on the ground-state problem, which can be exactly solved via the Bethe ansatz (we return to this below);
a recent, thorough review can be found in Ref.~\cite{BetheAnsatzReview}.

In this work we study the {\it thermodynamics} of unpolarized spin-$1/2$ fermions
with a contact interaction, i.e., the Gaudin-Yang model~\cite{GaudinYang}, 
\beq
\hat H = -\frac{\hbar^2}{2m}\sum_i \nabla_{i}^2 - \sum_{i < j} g\delta(x^{}_i-x^{}_j),
\eeq
where the sums are over all particles.
We cover weakly to strongly coupled regimes, as well as a wide range of temperatures,
and show lattice Monte Carlo results for the particle number density $n$, pressure $P$, compressibility $\kappa$, and
Tan's contact $\mathcal C$~\cite{Tan}. Furthermore, we use exact diagonalization on the lattice to obtain the second-order 
virial coefficient for the pressure $b^{}_2$, for which we also present analytic continuum results. Using the same analysis,
we obtain analytic and numerical answers for the leading-order coefficient for the contact $c^{}_2$.

\section{Many-body method, scales and dimensionless parameters} 
We employed a technique similar to that of Refs.~\cite{BDM1, BDM2, EoSUFG2} but applied in 1D.
The two-species fermion system is placed in a Euclidean space-time lattice of extent $N^{}_x \times N^{}_\tau$ with
periodic boundary conditions in the spatial direction and anti-periodic in the time direction. A Trotter-Suzuki decomposition 
of the Boltzmann weight is implemented, followed by a Hubbard-Stratonovich transformation, which allows us to write the grand-canonical 
partition function as a path integral over an auxiliary field. The path integral is evaluated using Metropolis-based Monte Carlo methods 
(see, e.g., Ref.~\cite{Drut:2012md}). 
Throughout this work, we use units such that $\hbar = m = k_B = 1$, where $m$ is the mass of the fermions.
The physical spatial extent of the lattice is $L = N^{}_x \ell$, and we take $\ell = 1$ to set the length and momentum scales.
The extent of the temporal lattice is set by the inverse temperature $\beta = 1/T = \tau N^{}_\tau$. The time step $\tau= 0.05$ 
(in lattice units) was chosen to balance temporal discretization effects with computational efficiency; in any case, those discretization 
effects are smaller than our statistical effects.

The physical input parameters are the inverse temperature $\beta$, the chemical potential $\mu = \mu^{}_\uparrow = \mu^{}_\downarrow$,
and the (attractive) coupling strength $g>0$. From these, we form two dimensionless 
quantities: the fugacity and the dimensionless coupling, given by 
\beq
z = \exp(\beta \mu)\ \ \ \ \ \text{and} \ \ \ \ \lambda^2 = \beta g^2, 
\eeq
respectively. In the grand-canonical ensemble, the density $n$ is an output variable, and therefore we
use $\lambda$ instead of the $\gamma = g/n$ parameter often employed in 1D ground-state studies 
(see, e.g., Refs.~\cite{Tokatly, FuchsRecatiZwerger}).

Note that 1D fermions with a contact interaction are 
ultraviolet-finite, and as a consequence the bare coupling has a physical meaning. In the continuum limit, $g = 2/a^{}_0$, where 
$a_0$ is the scattering length for the symmetric channel (see e.g. Ref.~\cite{ScatteringIn1D}). Using $z$ and $\lambda$ as parameters will
facilitate the comparison with experiments, as well as with other theoretical approaches.

Lattice calculations of the kind we use are exact, up to statistical and systematic uncertainties. To address the former, we have taken 5000 de-correlated samples for 
each data point in the plots shown below, which yields a statistical uncertainty of order $3\!-\!4\%$. To address the systematic effects, one must 
approach the continuum limit. Because one-dimensional problems are computationally inexpensive, it is possible to calculate 
in large lattices, from $N_x^{} = 50$ to 100 and beyond. For such lattice sizes, the continuum 
limit is achieved by lowering the density while still remaining in the many-particle, thermodynamic regime.  Operationally, this is 
accomplished by increasing the lattice parameter $\beta$, ensuring that the thermal wavelength $\lambda^{}_T = \sqrt{2 \pi \beta}$ 
satisfies $1 = \ell \ll \lambda^{}_T \ll L = \ell N^{}_x$; at fixed $z$, this reduces the density. 
In our calculations, we have used $\lambda^{}_T \simeq 3.5 - 7.0$ and $N^{}_x = 81$. 
We have then verified that our results collapse to the same 
(universal) curve when $\beta$ and $g$ are varied while $\lambda^2=\beta g^2$ is held fixed. This ``collapse'' takes place at different rates for different parameter
values (see Appendix \ref{App:Systematics} for additional details).
Lattice sizes larger than $N^{}_x = 81$ are computationally more expensive but certainly feasible; however, we chose to fix that size and cover 
a wider region of parameter space instead. Because our study proceeded at constant $\lambda$, increasing $\beta$ implies 
reducing $g$, which results in smaller uncertainties associated with the temporal lattice spacing $\tau$ in the 
Trotter-Suzuki decomposition; these are expected to be of order $1\!-\!2\%$ (see e.g. Ref.~\cite{BDM2}).  
\section{Results} 
We report our results in dimensionless form by displaying quantities in units of their non-interacting counterparts at the same 
value of the input parameters, or by scaling them by the appropriate power of the thermal wavelength $\lambda^{}_T = \sqrt{2 \pi \beta}$.
Among our results is the density equation of state $n(\lambda,\beta\mu)$, from which we obtain the pressure $P$ 
and the 
isothermal compressibility $\kappa$ by integrating and differentiating, respectively, with respect to the chemical potential.
Our last Monte Carlo result is Tan's contact $\mathcal C$, which we determine by computing the average interaction energy. In addition to these quantities, we
use exact diagonalization to compute the second-order virial coefficient for the pressure and density, and the corresponding 
leading-order coefficient for the contact; for both of these we also provide analytic results.

\subsection{Density} 
\begin{figure}[b]
\includegraphics[width=1.0\columnwidth]{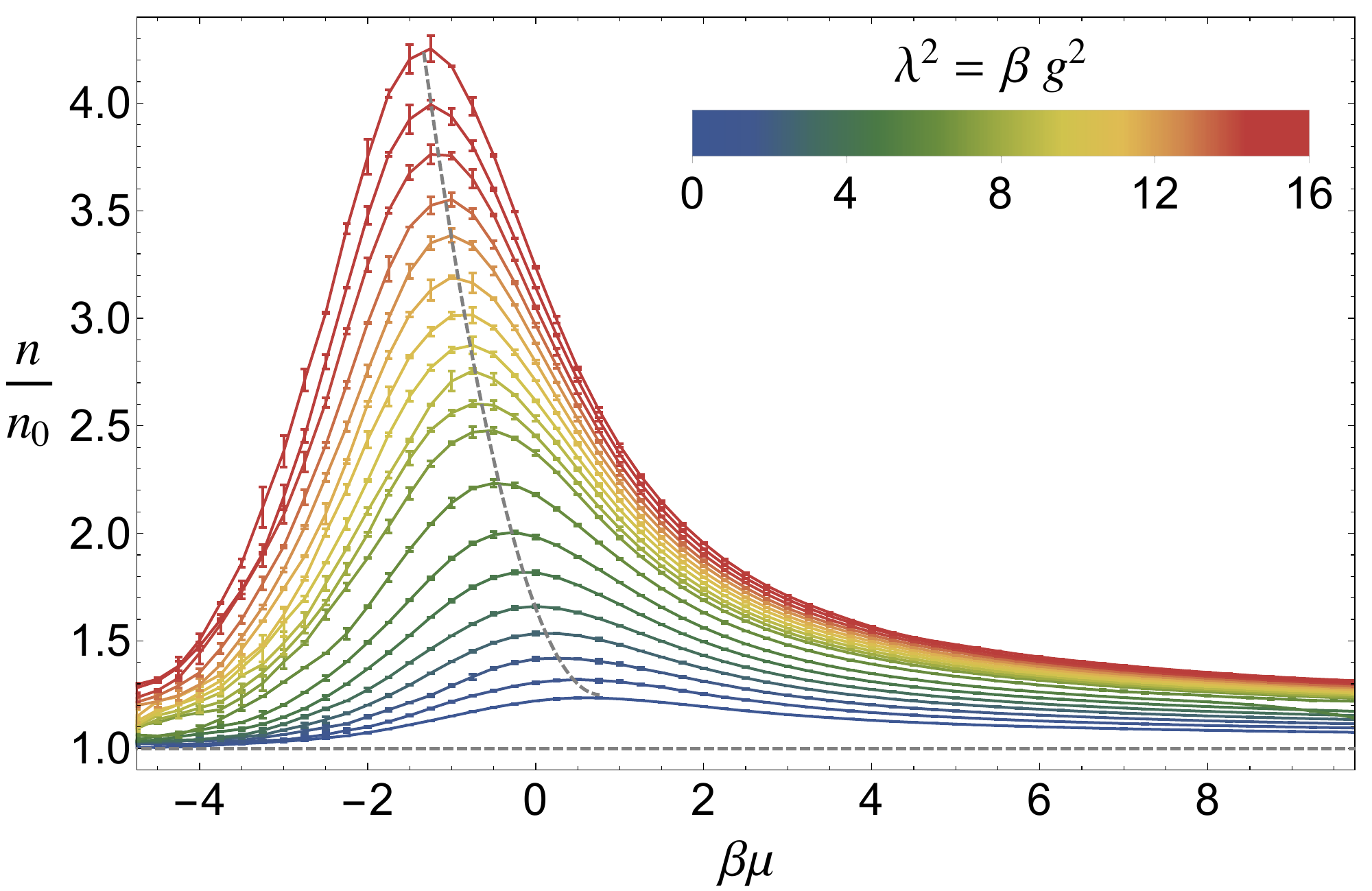}
\caption{\label{Fig:n_n0}(Color online) Density $n$, in units of the density of the non-interacting system $n^{}_0$, as a function
of the dimensionless parameters $\beta\mu\!=\!\ln z$ and $\lambda^2\!=\!\beta g^2$. 
From bottom to top, the coupling is $\lambda\!=\!0.0, 1.0, 1.25, 1.5, ... , 2.5, 2.75, 3.0, 3.1, 3.2, ... ,4.0$.
The dashed line joins the maxima at each $\lambda$.
}
\end{figure}
In Fig.~\ref{Fig:n_n0} we show the density $n$ as a function of the dimensionless parameters $z$ and $\lambda$,
defined above. The non-interacting result is
\beq
\label{Eq:n0}
n^{}_0 \lambda^{}_T = \frac{2}{\sqrt{\pi}} I^{}_1(z),
\eeq
where $I^{}_1(z) = z\,{d I^{}_0(z)}/{d z}$, and 
\beq
I^{}_0(z) = \int_{-\infty}^{\infty}dx \ln(1 + z e^{-x^2}).
\eeq
The solid curves in Fig.~\ref{Fig:n_n0} correspond to a three-point moving average over an interpolation of the 
original Monte Carlo data. The error bars represent the difference between the original data and the moving average.
For all $\lambda > 0$ there exists a strongly coupled regime around $\beta \mu=\ln z \simeq -1$,
where the deviation from the non-interacting system is maximal. This effect is more pronounced for larger $\lambda$. 
The locus of the maxima (indicated in Fig.~\ref{Fig:n_n0} with a dashed line) can be shown to satisfy $n^{}_0 \kappa^{}_0 = n \kappa$, 
where $\kappa$ is the isothermal compressibility of the system at finite $\lambda$ and $\kappa_0^{}$ is the noninteracting value.

These results are qualitatively similar to those of Ref.~\cite{Enss2D}. In that work, the density equation of state
was computed for the two-dimensional (2D) system. The similarity can be traced back to the fact that in both cases a bound state is
formed as soon as interactions are turned on, i.e. the unitary
limit coincides with the non-interacting limit. Therefore, increasing $\beta \mu$ along the line of constant physics (i.e. fixed $\lambda$) 
ultimately leads to a weak-coupling regime in 1D and 2D.
In three dimensions (3D), however, the analogous path drives the system deep into the non-trivial unitary limit. 
References~\cite{EoSUFG1,EoSUFG2}, for instance, do not see a peak in $n/n^{}_0$, but rather a monotonically increasing
function (see, e.g., Fig.~4(a) in Ref.~\cite{EoSUFG1}, or Fig.~4 in Ref.~\cite{EoSUFG2}). 

\begin{table}[b]
\begin{center}
\caption{\label{Table:FitParameters}
Fit parameters for the density equation of state, using the functional
form $n/n^{}_0 = 1 + \alpha ({\beta\mu})^{-\gamma}$, second-order pressure virial coefficient $b^{}_2$,
and leading-order contact virial coefficient $c^{}_2$, all as a function of the dimensionless coupling $\lambda$. 
For the non-interacting gas ($\lambda=0$), the virial coefficients are $b^{}_n = (-1)^{n+1} n^{-3/2}$.
}
\begin{tabularx}{\columnwidth}{@{\extracolsep{\fill}}c|c|c|c|c}
$\lambda$ & $b^{}_2 \text{(lattice)}$ & $c^{}_2 \text{(lattice)}$ & $\alpha$ & $\gamma$\\
\hline\hline
0 & $-0.35355...$  & $0$  & $0.0$ & $-$ \\ 
1.0 & $-0.035$  & $0.63$  & $0.24(1)$ & $0.46(6)$ \\
1.25 & $0.10$  & $1.26$  & $0.300(5)$ & $ 0.47(4)$\\
1.5 & $0.28$ & $2.36$  & $0.450(2)$ & $ 0.53(9)$\\
1.75 & $0.52$  & $4.22$  & $0.554(5)$ & $ 0.56(9)$\\
2.0 & $0.82$  & $7.33$  & $0.656(8)$ & $0.59(2)$\\
2.25 & $1.24$  & $12.5$  & $0.771(8)$ & $0.61(6)$\\
2.5 & $1.79$  & $21.0$  & $0.970(1)$ & $0.66(1)$\\
2.75 & $2.56$  & $34.9$  & $1.219(6)$ & $0.70(1)$\\
3.0 & $3.61$ & $57.7$  & $1.525(1)$ & $0.76(1)$\\
\hline\hline
\end{tabularx}
\end{center}
\end{table}

To characterize the approach to the non-interacting limit in the region $\beta\mu > 0$, we performed fits to the density using the 
(purely phenomenological) functional form
\beq
\label{Eq:FunctionFit}
{n}/{n^{}_0} = 1 + \alpha ({\beta\mu})^{-\gamma},
\eeq
where $\alpha,\gamma$ are functions of $\lambda$, as shown in Table~\ref{Table:FitParameters}.
For $\beta \mu \ll 0$, the virial expansion is applicable, for which
\beq
\label{Eq:VirialExpansion}
{n \lambda^{}_T}/{2} = z + 2 b_2^{} z^2 + 3 b_3^{} z^3 + \cdots ,
\eeq
and the factor of $1/2$ on the left-hand side comes from the number of fermion species.
In Table~\ref{Table:FitParameters} we show the virial coefficient $b^{}_2$ obtained by 
exact diagonalization of the two-body problem on the lattice. The exact result for $b^{}_2$
in the continuum limit, obtained by the same methods utilized in 3D (see e.g. Refs.~\cite{HoMueller, LeeSchaefer}),
is
\beq
\label{b2Exact}
b_2^{} = -\frac{1}{\sqrt{2}} + \frac{e^{\frac{\lambda^2}{4}}}{2\sqrt{2}}\left[1+\text{erf}(\lambda/2)\right],
\eeq
where $\text{erf}(x)$ is the error function.
From the above data, we determine other thermodynamic quantities, which 
furnish a prediction for ultracold atom experiments. 
%
\subsection{Temperature scale}
\begin{figure}[h]
\includegraphics[width=1.0\columnwidth]{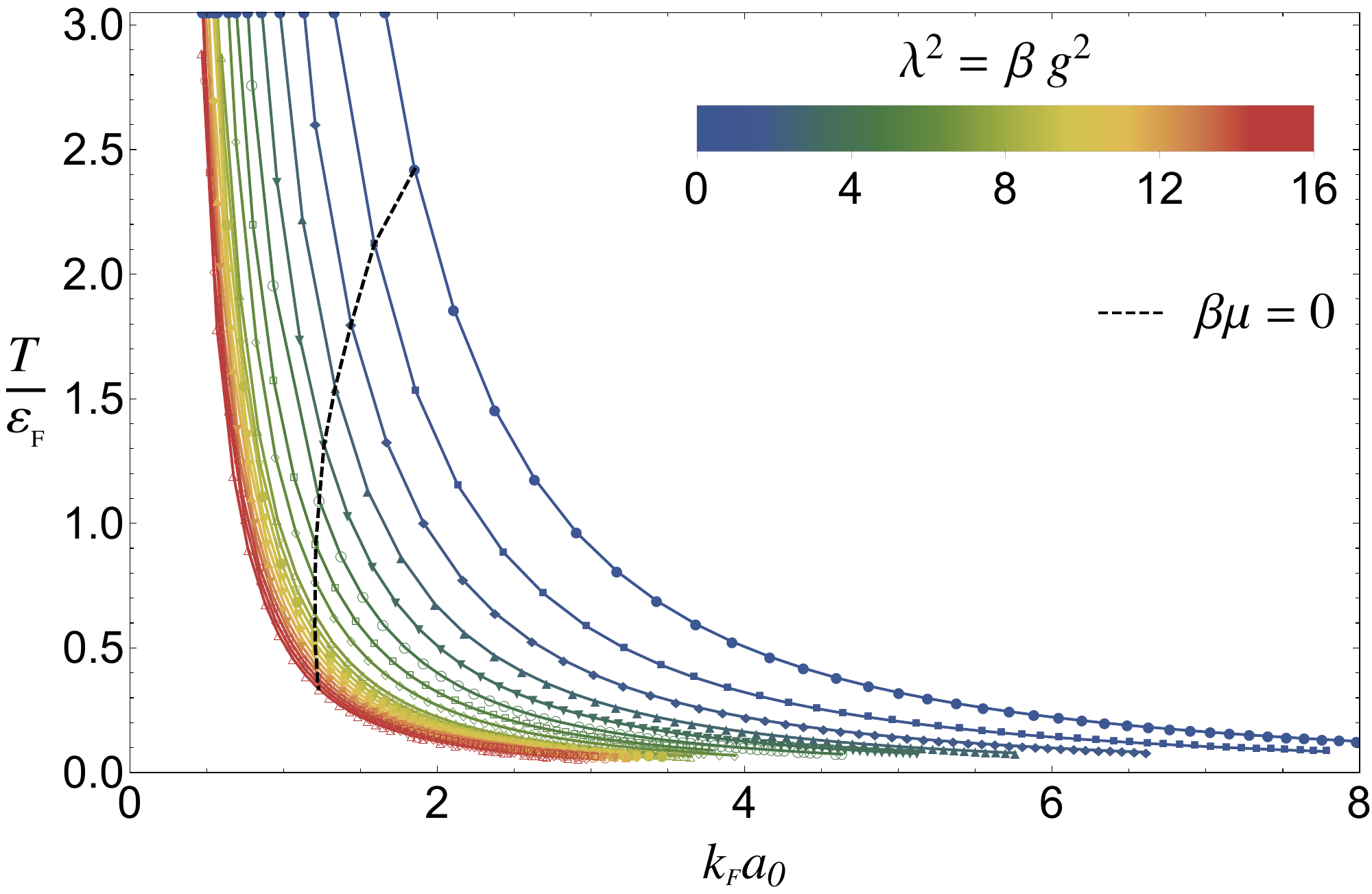}
\caption{\label{Fig:TVkF}(Color online) Temperature scale, in units of $\varepsilon_F^{}$, as a function of 
the coupling $k_F^{}a^{}_0$. Here, $k_F^{} = \pi n/2$, where $n$ is the total density, and $\varepsilon_F^{}=k_F^{2}/2$.
The dashed line connects the $\beta \mu=0$ points for each value of $\lambda$. The $\beta \mu > 0$ ($<0$) points lie to the 
right (left) of the dashed line.
}
\end{figure}

Having the density as a function of $\beta \mu$ at our disposal, we can determine the temperature
scale in a different convention which is often used, namely $T/\varepsilon^{}_F$, where $\varepsilon^{}_F=k_F^{2}/2$
and $k_F^{} = \pi n/2$. In Fig.~\ref{Fig:TVkF} we show our results for $T/\varepsilon{}_F$
as a function of the dimensionless coupling $k_F^{}a_0^{}$, for each value of $\lambda$. This graph should be 
understood as a parametric plot: both axes depend on $\beta \mu$ implicitly through $n$, at fixed $\lambda$.
As can be appreciated from this plot, for each $\lambda$ our results cover a range in $T/\varepsilon^{}_F$ that
goes from below $0.1$ all the way to beyond $1.5$ (for display purposes, Fig.~\ref{Fig:TVkF} does not show the full 
upper region of the $T/\varepsilon^{}_F$ axis, which corresponds to large, negative $\beta \mu$).

\subsection{Pressure and compressibility} 
\begin{figure}[h]
\includegraphics[width=1.0\columnwidth]{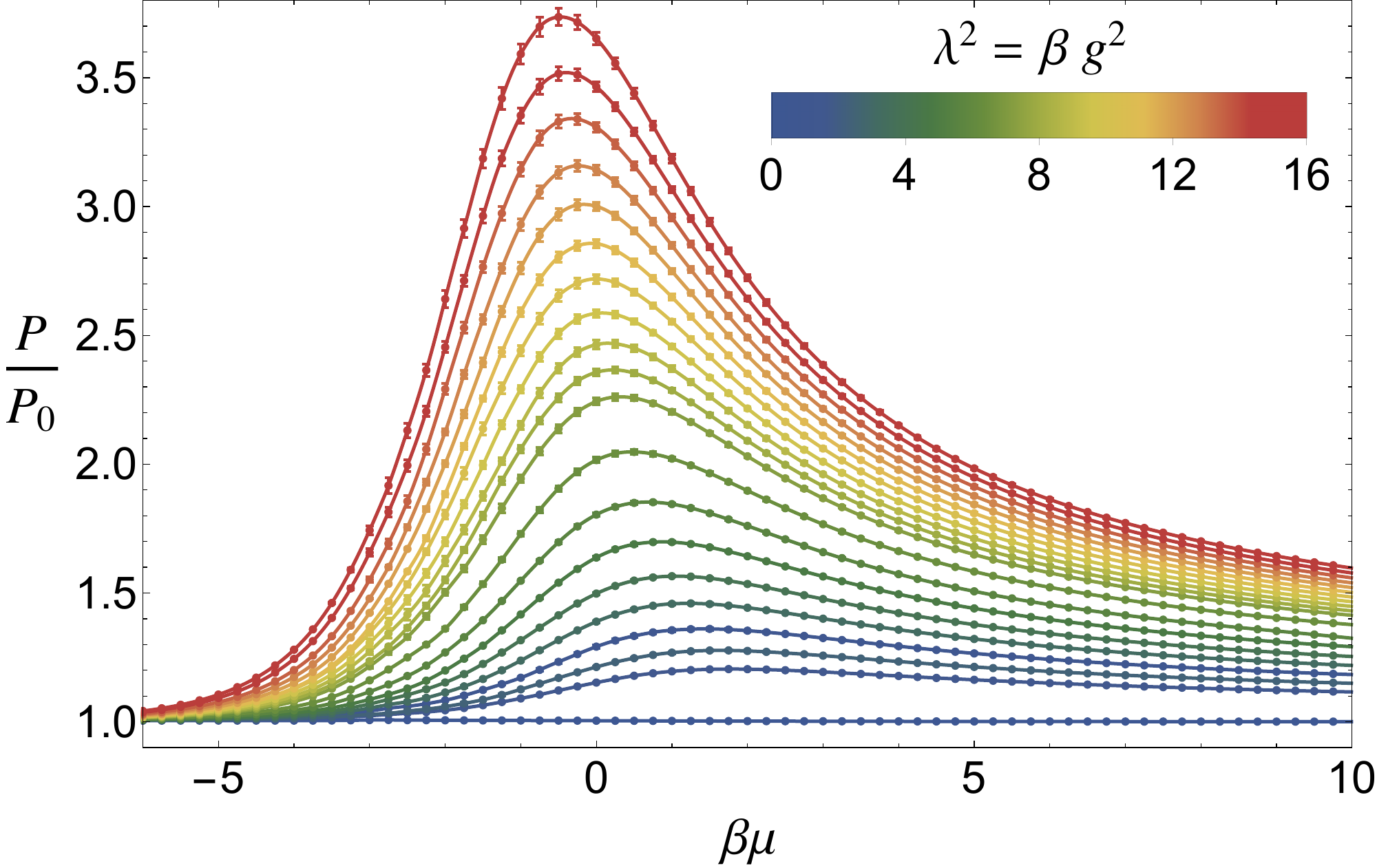}
\caption{\label{Fig:P_P0}(Color online) Pressure in units of its non-interacting counterpart, as a function
of the dimensionless parameters $\beta \mu = \ln z$ and $\lambda^2 = \beta g^2$, obtained by 
$\beta\mu$-integration of the density (see Eq.~\ref{Eq:PIntegration}). The values of $\lambda$ shown in this plot
are the same as in Fig.~\ref{Fig:n_n0}.}
\end{figure}

\begin{figure}[h]
\includegraphics[width=1.0\columnwidth]{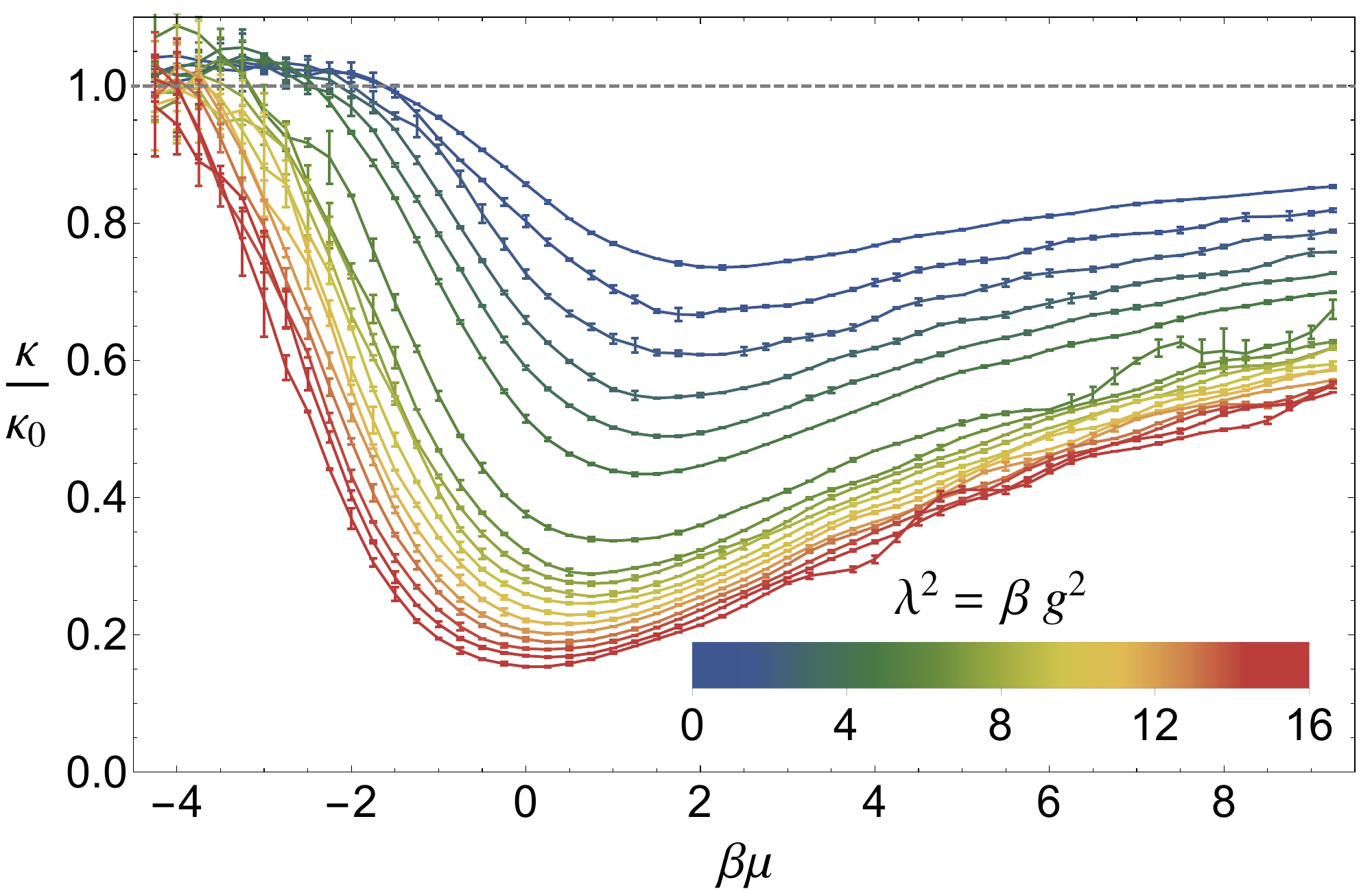}
\caption{\label{Fig:kappa}(Color online) Isothermal compressibility in units of its non-interacting counterpart, as a function
of the dimensionless parameters $\beta \mu = \ln z$ and $\lambda^2 = \beta g^2$. The values of $\lambda$ shown in this 
plot are the same as in Fig.~\ref{Fig:n_n0}, but from top to bottom instead.}
\end{figure}

It is straightforward to obtain an estimate for the pressure by integrating $n \lambda^{}_T$
over $\log z = \beta \mu$. We take the $z=0$ limit (i.e., $\beta \mu \to -\infty$) as a reference point. In practice, we
verify that the data heals (within statistical uncertainties) to the virial expansion at low $z$, and use
that result (at second order) to complete the integration to $z=0$. In that limit the pressure vanishes, such that
\beq
\label{Eq:PIntegration}
P\lambda_T^3 = 2\pi \int_{-\infty}^{\beta \mu} {n \lambda_T} \; d (\beta \mu)'.
\eeq
The results for $P$, in units of the non-interacting pressure $P_0$, are shown in Fig.~\ref{Fig:P_P0}. Note that
%
$P^{}_0\lambda_T^3 = \sqrt{16 \pi} I^{}_0(z)$,
%
where $I^{}_0(z)$ is given above.
By taking a derivative of $n$ one obtains the isothermal compressibility,
\beq
\kappa = \frac{\beta}{n^2}\left . \frac{\partial n}{\partial (\beta \mu)} \right |^{}_\beta = 
\lambda^{3}_T \frac{\sqrt{2\pi}}{(n \lambda^{}_T)^2} \left . \frac{\partial (n \lambda^{}_T)}{\partial (\beta \mu)} \right |^{}_\beta .
\eeq
We report this quantity in Fig.~\ref{Fig:kappa}, in units of its non-interacting counterpart $\kappa^{}_0$, where
(in dimensionless form) $\kappa^{}_0 \lambda_T^{-3} = \pi^{-3/2} (n^{}_0 \lambda^{}_T)^{-2} I_2(z)$,
and $I^{}_2(z) = z\,{d I^{}_1(z)}/{d z}$. As expected, in the limits of large $\beta \mu$ (both positive and negative) $\kappa$ 
tends to $\kappa^{}_0$. On the other hand, in the strongly interacting region $\kappa \ll \kappa^{}_0$, i.e., the system is 
less compressible than in the non-interacting regime. We attribute this to the formation of localized di-fermion molecules
and Pauli exclusion.  Note that oscillations in these curves at large $\beta \mu$ reflect the inherent  instability of  calculating numerical derivatives (when coupled with the statistical uncertainty in $n$), rather than a physical effect.

\subsection{Contact} 
Knowing the density as detailed above, one may use the Maxwell relation to calculate the contact $\mathcal C$ 
from $n$ (see Refs.~\cite{ThermoContact,MaxwellContact1D,Tan}),
which in dimensionless form reads
\beq
z \left . \frac{\partial (\beta^2 \mathcal C)}{\partial z} \right |^{}_{\lambda,T}=  \frac{\lambda^2}{2 \sqrt{2\pi}} \left . \frac{\partial (n \lambda^{}_T )}{\partial \lambda} \right |^{}_{z,T}.
\eeq
Alternatively, one may use the interaction energy $\langle \hat V \rangle$. Starting from the definition in 1D,
\beq
{\mathcal C} = \frac{2}{\beta \lambda^{}_T}\left . \frac{\partial (\beta \Omega)}{\partial (a^{}_{0}/\lambda^{}_T)} \right |^{}_{\mu,T},
\eeq
where $\Omega$ is the grand thermodynamic potential, the contact can be shown, using the Feynman-Hellman theorem, to be
given by
\beq
{\mathcal C} = -{g} {\langle \hat V \rangle}. 
\eeq
Note that ${\mathcal C}$ can be made dimensionless and intensive by multiplying it by $\lambda^{4}_T/L$.
On the other hand, the virial expansion for $\Omega$ reads 
%
$
- \beta \Omega =  Q_1^{} \left(z + b^{}_2 z^2 + b^{}_3 z^3 + \dots \right),
$
%
where $Q_1^{} = 2 L / \lambda^{}_T$ is the single-particle partition function, and the virial coefficients $b_n^{}$ are the same as those for 
the density appearing in Eq.~\ref{Eq:VirialExpansion}. Thus, the virial expansion for $\mathcal C$ takes the form
\beq
\label{Eq:BetaContact}
\beta {\mathcal C} = \frac{2}{\lambda^{}_T}  Q_1^{} \left(c^{}_2 z^2 + c^{}_3 z^3 + \dots \right),
\eeq
where 
\beq
\label{Eq:Cn}
c^{}_n = -\frac{\partial b_n^{}}{\partial \left(a^{}_0/\lambda^{}_T \right)}
= \sqrt{\frac{\pi}{2}} \lambda^2 \frac{\partial b_n^{}}{\partial \lambda}.
\eeq
Our definition for the $c_n^{}$ coefficients coincides with that of Ref.~\cite{VignoloMinguzzi}. 
From our calculation of the virial coefficient $b^{}_2$, we obtain ${\partial b_2^{}}/{\partial \lambda}$;
the resulting $c_2^{}$ is shown in Table~\ref{Table:FitParameters}.
The exact continuum result (based on Eqs.~\ref{b2Exact} and~\ref{Eq:Cn}) is
\beq
\label{c2Exact}
c_2^{} = \frac{\lambda^2}{4} + \frac{\sqrt{\pi}}{8}e^{\frac{\lambda^2}{4}}\lambda^3\left[1+\text{erf}(\lambda/2)\right].
\eeq

\begin{figure}[t]
\includegraphics[width=1.0\columnwidth]{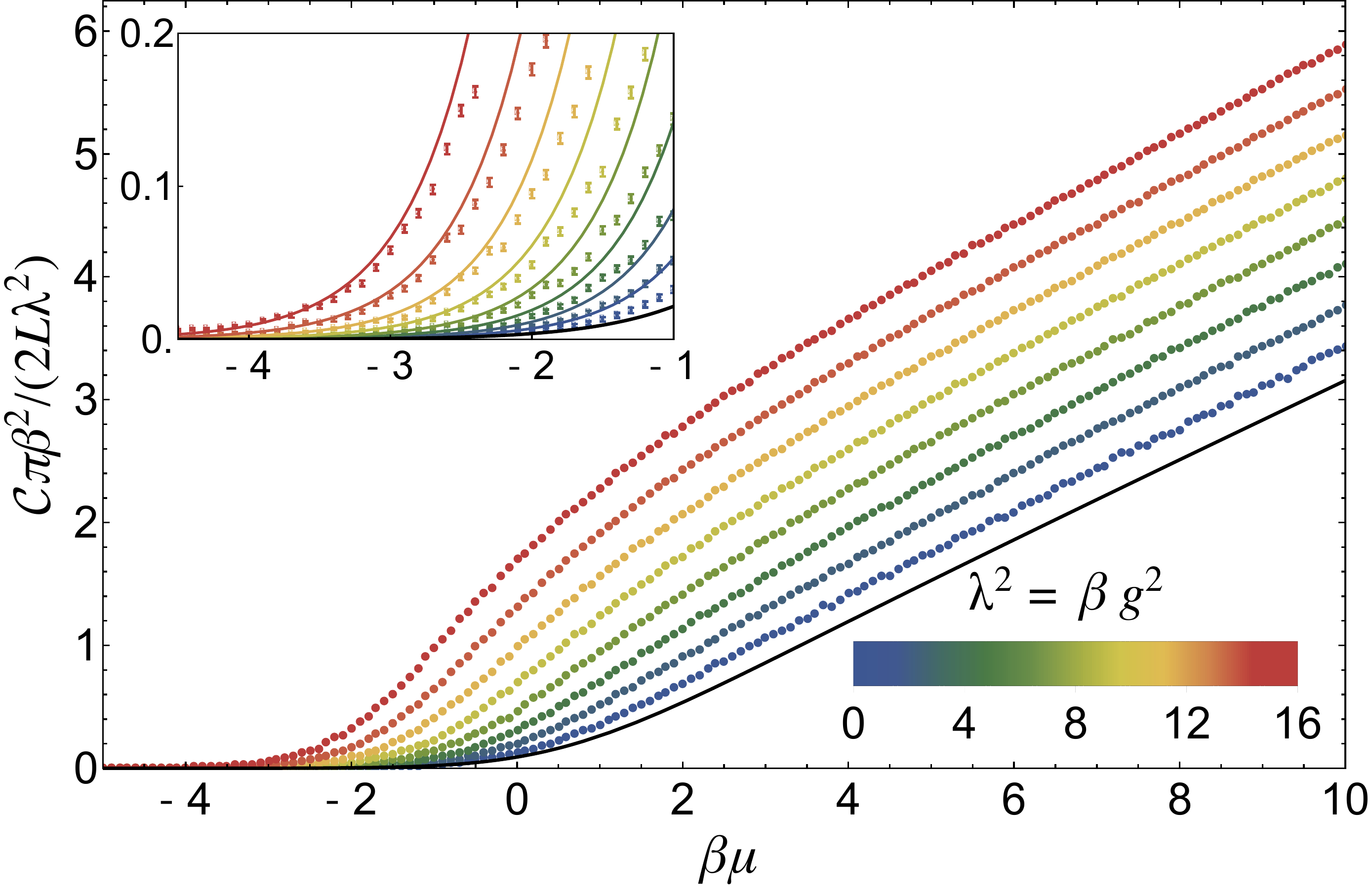}
\caption{\label{Fig:Contact}(Color online) Tan's contact $\mathcal C$, scaled by 
$\beta \lambda^{}_T/(2 Q^{}_1 \lambda^2 )  = \pi \beta^2/(2 L \lambda^2 )$ [see Eqs.~(\ref{Eq:BetaContact}) and (\ref{Eq:Cn})], as a function
of $\beta \mu$. The black line shows $\mathcal C$ in the absence of interactions. Inset: Zoom-in of main plot on the region $-4.5 \leq \beta\mu \leq -1.0$, showing also the leading-order virial 
expansion. Both plots show data for $\lambda = 0.5,1.0,1.5,...,4.0$, which appear from bottom to top.}
\end{figure}

In Fig.~\ref{Fig:Contact} we show our results for the contact, including the leading-order virial expansion (inset).
We show statistical error bars in the inset; in the main plot, the smoothness of the results across $\beta \mu$
indicate that the statistical effects are of the order of the size of the symbols.
As seen in the inset, the data captures the correct asymptotic behavior at small $z$ for all $\lambda$, but
the agreement slowly deteriorates at large $\lambda$, suggesting that the virial expansion breaks down earlier in
that regime. For $\beta \mu \gg 1$ the contact satisfies
\beq
\mathcal C \pi \beta^2/(2 L \lambda^2 ) = \langle \hat n^{}_\downarrow \hat n^{}_\uparrow \rangle \pi \beta/2 \to \zeta^{}_1 \beta \mu + \zeta^{}_2,
\eeq
where we find $\zeta^{}_1 = 0.35(1)$ is nearly $\lambda$-independent; 
it is in fact a feature of density-density correlations in the non-interacting gas that 
leaves an imprint at all couplings (see below). On the other hand, as is evident from the plot, $\zeta^{}_2$ is approximately linear 
in $\lambda$ at large $\beta \mu$ [we find $\zeta^{}_2(\lambda) \simeq a + b \lambda$ with $a = -0.34(1)$ 
and $b = 0.701(5)$ at $\beta\mu=10$]. Analytic estimates in the absence of interactions yield 
$\zeta^{}_1 = 1/\pi = 0.318...$ and $\zeta^{}_2 \propto (\beta\mu)^{-1}$.
Although much is known about $\mathcal C$ in various situations (see e.g. Ref.~\cite{ContactReview} for a review),
the full temperature dependence in 1D shown here does not appear anywhere else in the literature, to the best of our knowledge.
%

\section{Summary and Conclusions} 
We have performed a controlled, fully non-perturbative study of the 
thermodynamics of the Gaudin-Yang model (i.e., a one-dimensional, two-species Fermi system, with short-range, attractive
interactions). We employed lattice Monte Carlo methods that have been successfully utilized before for similar studies, 
and discussed statistical and systematic uncertainties. We report here on several quantities, namely
the density, pressure, contact, and leading virial coefficients, in all cases covering weakly to strongly coupled regimes 
(as characterized by values of the dimensionless parameter $0 \leq \lambda^2 \leq16.0$), as well
as low to high temperatures (as characterized by $-5.0 \leq \beta \mu \leq 8.0$, which ranges from the semi-classical 
regime $\beta \mu < -1.0$ to the deep quantum regime $\beta \mu > 1.0$, which we also display in terms of $T/\varepsilon^{}_F$).
Our results for the density equation of state display a behavior similar to that observed in 2D systems: A regime 
exists around $\beta \mu = \ln z \simeq -1$ in which deviations from the non-interacting case are maximal. As $z$ is increased from 
$z\ll1$ (the semi-classical regime where the virial expansion is valid) this strongly coupled regime is (roughly) accompanied by 
the onset of quantum fluctuations at $\beta \mu = \ln z \simeq 0$.

Although certain 1D Fermi systems are exactly solvable via the Bethe ansatz~\cite{BatchelorEtAl,TakahashiBook}, the latter is restricted to 
uniform systems in the ground state (or close to it~\cite{BatchelorEtAlGaudinYang}). Indeed, finite temperature studies require the 
thermodynamic Bethe ansatz, which involves solving an infinite tower of coupled non-linear integral 
equations~\cite{BatchelorEtAlGaudinYang}.  The necessary truncation of this tower  leads to a potentially uncontrolled approximation, in contrast to the control over uncertainties present in the Monte Carlo techniques used here. 
Regardless, it is somewhat surprising that a thorough numerical characterization of this simple system, as a benchmark for 
many-body methods, is absent from the literature, to the best of our knowledge. To help remedy this situation as much as possible, we have
aimed to characterize the universal thermodynamics of this system in detail.

Finally, our results constitute predictions for experiments with ultracold atoms in highly elongated optical traps. These are
now realized using modulated potentials. Moreover, our results are {\it universal} in the sense that they apply to any
unpolarized atomic gas in dilute regimes, where the interaction potential is well approximated by a contact interaction.
As we show throughout all reported quantities, there is only one interaction parameter (i.e. $\lambda$) determining the 
thermodynamics.
Our study is readily generalizable to a higher number of fermion species, which are expected to be experimentally available 
in the near future~\cite{LargeNf}.

\acknowledgments 
This material is based upon work supported by the National Science Foundation 
Graduate Research Fellowship Program under Grant No. DGE{1144081}, National Science Foundation Nuclear Theory Program
under Grant No. PHY{1306520} and National Science Foundation REU Sites Program under Grant No. ACI{1156614}.

\appendix
\section{Systematics of the approach to the continuum limit \label{App:Systematics}}

\begin{figure}[t]
\includegraphics[width=1.0\columnwidth]{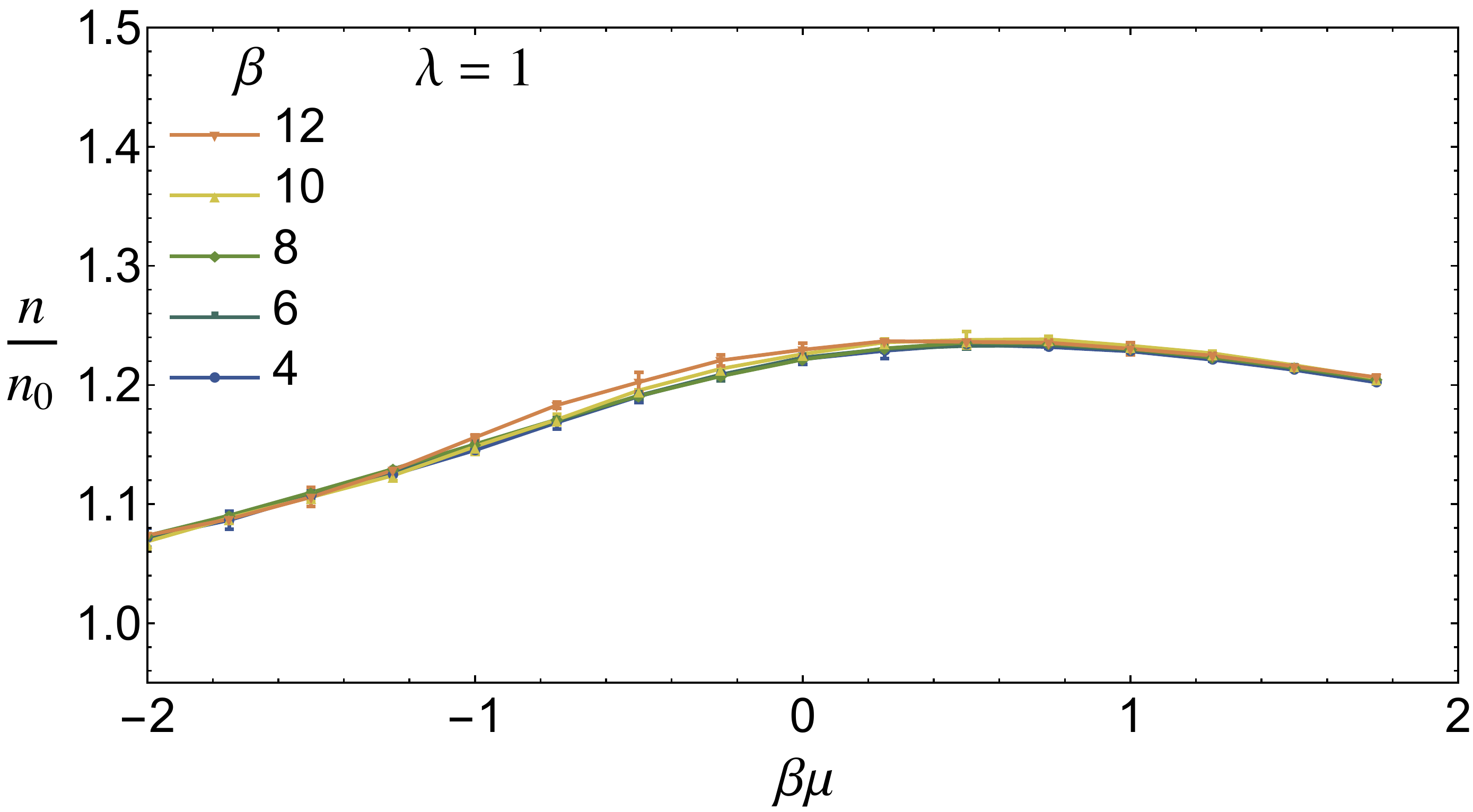}
\caption{\label{Fig:BetaSystematicsWC}(Color online) Density $n$, in units of the non-interacting density $n_0^{}$,
as a function of $\beta\mu$ at weak coupling ($\lambda = 1.0$), for several values of $\beta$. Finite-$\beta$ effects are small
throughout the graph. Note the ranges in the $x$ and $y$ axes are different from those of Fig.~\ref{Fig:n_n0}.}
\end{figure}

\begin{figure}[b]
\includegraphics[width=1.0\columnwidth]{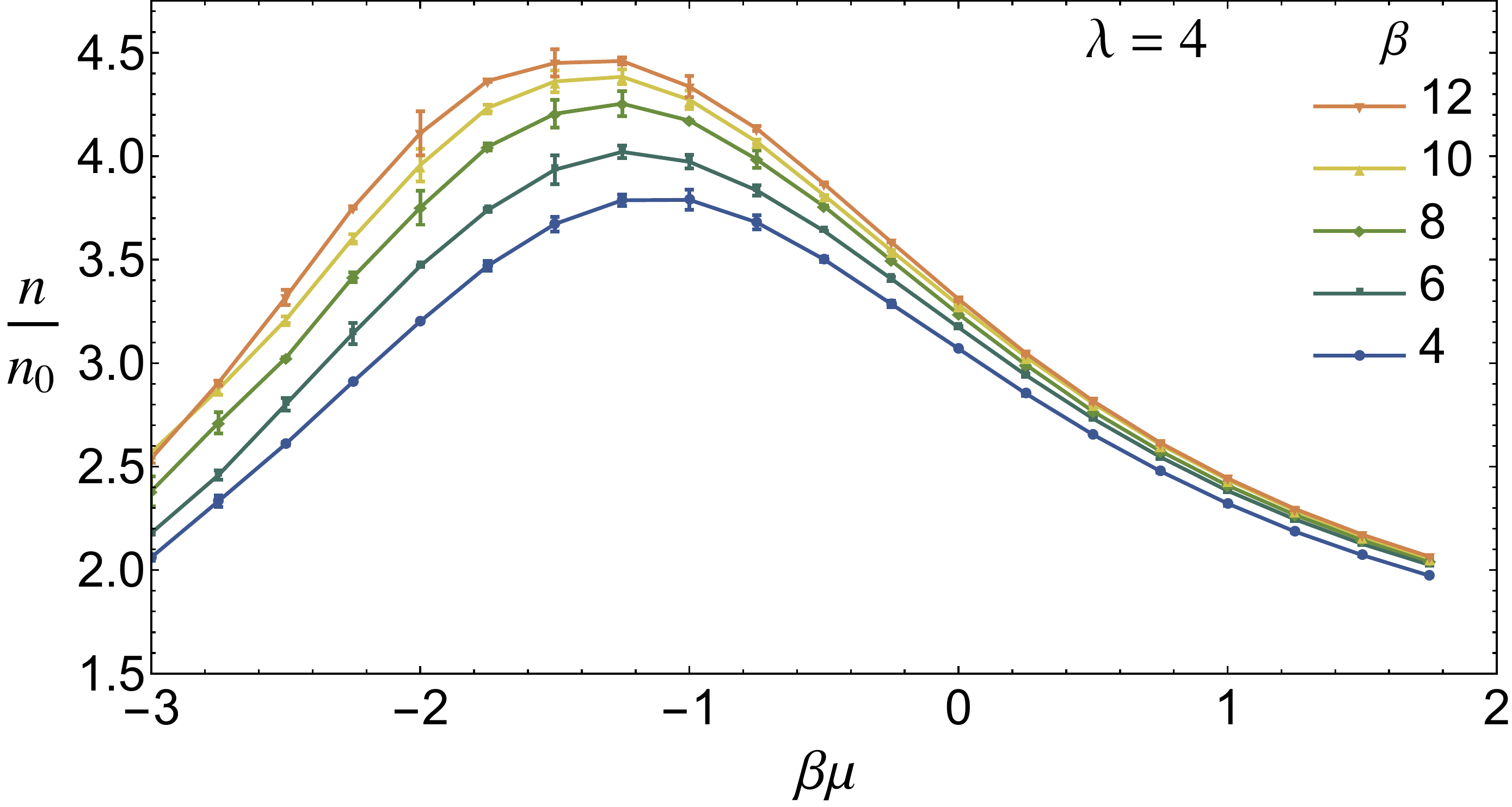}
\caption{\label{Fig:BetaSystematicsSC}(Color online) Density $n$, in units of the non-interacting density $n_0^{}$,
as a function of $\beta\mu$ at the strongest coupling in this study ($\lambda = 4.0$), for several values of $\beta$. Finite-$\beta$ effects are clearly 
visible, especially around the maximum. Note that the $x$-axis range is different from that of Fig.~\ref{Fig:n_n0}, but the $y$-axis is slightly extended.}
\end{figure}

In this section we report briefly on the systematic effects resulting from performing calculations at finite $\beta$.
As mentioned in the main text, the continuum limit is approached in our method when $\beta \to \infty$, and
different quantities approach their limit at different rates, which also depend on the values of other input parameters
(e.g. $\beta \mu$). As we show in Figs.~\ref{Fig:BetaSystematicsWC} and~\ref{Fig:BetaSystematicsSC}, the convergence to the large-$\beta$
limit improves as the difference between  $\beta\mu$ and the  $\beta\mu\simeq-1$ point (where the interaction and quantum effects dominate)  increases. This is clearer at
strong coupling (Fig.~\ref{Fig:BetaSystematicsSC}) than at weak coupling (Fig.~\ref{Fig:BetaSystematicsWC}); indeed,
the latter is essentially converged already at $\beta = 4$, whereas the former still shows finite-$\beta$ effects
even at $\beta=8$ in some regions. From these graphs, we infer that the largest systematic uncertainties 
due to finite $\beta$ are on the order of $10\%$. We stress that that is an upper bound for these systematic effects.
Those effects are most prominent around the maximum in $n/n^{}_0$;
they are apparent for the strongest couplings we have studied ($\lambda=4$) and are small for weak coupling ($\lambda=1$).
%

\end{document}